\documentclass[12pt]{article}
\usepackage{epsfig}
\usepackage{amsmath}
\usepackage{hhline}
\usepackage{amssymb}
\usepackage{times}
\usepackage{color}
\usepackage{cite}
\usepackage{url}

\newlength{\dinwidth}
\newlength{\dinmargin}
\definecolor{orange}{rgb}{1,0.8,0.3}                     

\usepackage[tight]{subfigure}

\newcommand*{\br}[1] {\left( #1 \right)}				
\DeclareMathOperator{\erf}{erf}						
\newcommand*{\Eslash}{``E\kern-0.4em\slash$\,$''}			
			
\newcommand*{\noEslash}{``no~E\kern-0.4em\slash$\,$''}			
\DeclareMathOperator{\sign}{sign}					
					
\DeclareMathOperator{\GeV}{GeV}					
\DeclareMathOperator{\TeV}{TeV}					
\DeclareMathOperator{\der}{d}

\newcommand{\artitle}[1]{}
\newcommand{\artitlekeep}[1]{``#1''}
\newcommand{\sub}[1]{\ensuremath{_{\mathrm{#1}}}} 
\newcommand{\super}[1]{\ensuremath{^{\mathrm{#1}}}} 
\newcommand{\mWgen}{\ensuremath{m\sub{W}\super{gen}=80.419\,\GeV}}

\newcommand{\Egamma}{\ensuremath{E\sub{\gamma}}}			
\newcommand{\Egammagen}{\ensuremath{E\sub{\gamma}\super{gen}}}		
\newcommand{\pz}{\ensuremath{p\sub{z}}}					
\newcommand{\pxgamma}{\ensuremath{p\sub{x,\gamma}}}			
\newcommand{\pygamma}{\ensuremath{p\sub{y,\gamma}}}			
\newcommand{\pzgamma}{\ensuremath{p\sub{z,\gamma}}}			
\newcommand{\pzgammagen}{\ensuremath{p\sub{z,\gamma}\super{gen}}}	
\newcommand{\Ebeam}{\ensuremath{E\sub{beam}}}				
\newcommand{\EISR}{\ensuremath{E\sub{ISR}}}				
\newcommand{\Emax}{\ensuremath{E\sub{max}}}				
\newcommand{\E}{{\mathrm{e}}}

\begin{document}

\newcommand {\gapprox}
   {\raisebox{-0.7ex}{$\stackrel {\textstyle>}{\sim}$}}
\newcommand {\lapprox}
   {\raisebox{-0.7ex}{$\stackrel {\textstyle<}{\sim}$}}
\def\gsim{\,\lower.25ex\hbox{$\scriptstyle\sim$}\kern-1.30ex%
\raise 0.55ex\hbox{$\scriptstyle >$}\,}
\def\lsim{\,\lower.25ex\hbox{$\scriptstyle\sim$}\kern-1.30ex%
\raise 0.55ex\hbox{$\scriptstyle <$}\,}

%
%

\begin{titlepage}
\begin{flushleft}
{\tt DESY 10-078    \hfill    ISSN 0418-9833} \\
{\tt May 2010}                  \\
\end{flushleft}

\vspace{1.0cm}

\begin{center}
\begin{Large}

{\bfseries \boldmath Treatment of Photon Radiation in Kinematic Fits \\
at Future $e^+ e^-$ Colliders}

\vspace{1.5cm}

M.~Beckmann$^{1}$, B.~List$^{2}$, and J.~List$^{1}$

\end{Large}

\vspace{.3cm}
1- Deutsches Elektronen-Synchrotron DESY\\ 
   Notkestr. 85 \\
   22607 Hamburg, Germany
\vspace{.1cm}\\
2- University of Hamburg\\
   Institute for Experimental Physics\\
   Luruper Chaussee 149\\
   22761 Hamburg, Germany

\end{center}

\vspace{1cm}

\begin{abstract}
Kinematic fitting, where constraints such as energy and momentum 
conservation are imposed on measured four-vectors of jets and leptons,
is an important tool to improve the resolution in high-energy 
physics experiments. 
At future $e^+ e^-$ colliders, photon radiation parallel to the beam
carrying away large amounts of energy and momentum will become a 
challenge for kinematic fitting. 
A photon with longitudinal momentum 
$\pzgamma\,(\eta)$ is introduced, which is parametrized such that
$\eta$ follows a normal distribution.
In the fit, $\eta$ is treated as having a measured value of zero,
which corresponds to $\pzgamma=0$.
As a result,
fits with constraints on energy and momentum conservation
converge well even in the presence of a highly energetic photon,
while the resolution of fits without such a photon
is retained.
A fully simulated and reconstructed 
$e^+ e^- \rightarrow q \bar q q \bar q$ 
event sample at $\sqrt{s}=500\,\GeV$
is used to investigate the performance of this method
under realistic conditions,
as expected at the International Linear Collider.
\end{abstract}

\vspace{1.0cm}

\begin{center}
To be submitted to {Nucl. Instrum. Meth.} {\bf A}
\end{center}

\end{titlepage}


\section{Introduction}
\label{sec:intro}

Radiation of photons at angles so small that they escape along 
the beam pipe is usually not taken into account in kinematic fits. 
At previous $e^+ e^-$ colliders such as LEP, the losses due to photon 
radiation were acceptable \cite{lep}. At future facilities 
such as the International Linear Collider (ILC) \cite{bib:ilc} or 
the Compact Linear Collider (CLIC) \cite{bib:clic}, photon 
radiation will be much stronger due to higher center-of-mass energies and 
stronger focussing of the beams,
which makes it desirable to model photon radiation in kinematic fits.

Kinematic fitting is a well-established tool to improve jet energy and 
invariant mass resolutions. A number of four-vectors 
representing the final state particles is fitted under 
constraints such as energy and momentum conservation. 
The four-vectors are parametrized by suitably chosen variables
such that the measured values 
follow an approximately Gaussian distribution
around the true values.
A $\chi^2$ that quantifies the deviation between 
measured and fitted parameters is minimized under the condition that 
the imposed constraints are fulfilled \cite{bib:kinfitnote}.

The improvement in resolution emerges from the redundant information 
contained in the measured values in the presence of constraints. 
Unmeasured parameters reduce the redundancy, 
since one constraint is used up for each unmeasured parameter to 
determine its value. The redundancy is quantified by the number of 
degrees of freedom, which is given by the number of constraints minus 
the number of unmeasured parameters.

The two main effects that cause the emission of photons collinear with 
the incoming beams so that they escape the main detector are initial state 
radiation (ISR) and beamstrahlung. ISR is a higher-order QED effect, at 
which real photons are emitted before the actual interaction. 
Beamstrahlung is caused by the electrical fields of the bunches colliding 
with each other: electrons in the one bunch are deflected by the field of 
the other bunch and thus emit bremsstrahlung photons.

ISR is characterized by an energy spectrum that follows a power law with 
an exponent of roughly $-0.9$ \cite{bib:isr}. Thus the vast majority of events have at most
one ISR photon with an energy above a few GeV, which is the accuracy to which
the total energy and longitudinal momentum of fully hadronic events
can be measured by a typical detector envisioned for the linear collider.
This photon can, however, carry substantial
energy of tens of GeV.
Beamstrahlung on the other hand has an energy spectrum with an
exponent of $-2/3$, but with an additional exponential suppression of high
energy photons \cite{bib:bs}. The mean number of beamstrahlung photons
emitted prior to the interaction can be of order one or even larger, depending
on the beam parameters.

This paper presents a novel method to take the energy and longitudinal
momentum of photon radiation into account in kinematic fits. 
A priori information 
about the momentum spectrum of photon radiation is used to 
treat the photon's momentum as a measured parameter in the fit. 
As a test case, the production of $W^{+}W^{-}$/$Z^0Z^0$ pairs
decaying to light quark jets at the ILC is considered,
with fully simulated Monte Carlo events as reconstructed by the
International Large Detector (ILD) \cite{bib:ild}. 
A more detailed description of the method and its application tests 
can be found in \cite{bib:diplomarbeit}.

The main focus of this method is an improved
treatment of the effects of ISR, because ISR is the main source of highly energetic
photons. Therefore, only a single photon is included in the kinematic
fit, with an energy spectrum given by a power law, as expected for ISR. 
A similar method with the inclusion of two photons in the fit and an
energy spectrum describing the combined effects of ISR and beamstrahlung is the
subject of ongoing work and beyond the scope of the current publication.
However, the method presented here leads to a significant improvement also
in the presence of beamstrahlung, as shown in section~\ref{sec:performance}.


\section{Representation of the photon}
\label{sec:representation}

Since photons from ISR and beamstrahlung escaping the detector
have to a good approximation
zero transverse momentum with respect to the beam direction,
they affect mainly the conservation of (detected) energy $E$ and longitudinal 
momentum $\pz$.
The simplest method to cope with highly energetic photons in a constrained
kinematic fit is therefore to drop the energy and longitudinal momentum conservation 
constraints,
thus losing two degress of freedom.

A somewhat better solution is to introduce a fit object representing
the undetected photon with one free, unmeasured parameter, namely
its longitudinal momentum $\pzgamma$, and 
set $\pxgamma=\pygamma=0$ and thus $\Egamma=|\pzgamma|$.
This allows the energy and $\pz$ constraints to be recovered,
at the price of one unmeasured parameter, so that one degree of freedom
is regained.

However, this approach neglects the information about the momentum 
spectrum of the photons. 
Here this information is used so that the photon is treated as
a particle with a measured momentum of zero and an uncertainty 
derived from its known momentum spectrum.

\subsection{Parametrization of the Photon Energy}
\label{subsec:parametrization}

In a kinematic $\chi^2$ fit the measured four-vector components
of a particle or jet are parametrized with parameters $\eta\sub{i}$ (e.g.,
$E, \theta, \phi$)
such that the difference $\eta\sub{i, meas} - \eta\sub{i, true}$
between the measured $\eta\sub{i, meas}$ and the true value 
$\eta\sub{i, true}$ follows a Gaussian distribution with zero mean and 
standard deviation $\delta \eta\sub{i}$
(for reasons of notational simplicity we limit the discussion to
the case where the parameters $\eta\sub{i}$ are uncorrelated).
Then,
\begin{equation}
  \chi^2 = \sum \limits_{i} \frac{(\eta\sub{i, meas} - \eta\sub{i})^2}
  {\delta \eta\sub{i}^2}
\end{equation}
 is, apart
from a constant,
identical to the negative logarithm of the likelihood to obtain the measured
values, given the values $\eta\sub{i}$:
\begin{equation}
  \chi^2 = -\ln {\mathcal {P}}\,(\eta\sub{i, meas} | \eta\sub{i}) +
  const.
\end{equation}
Thus, the $\chi^2$ fit seeks the best estimate $\eta\sub{i}$
for the true parameter values by maximizing the likelihood 
to get the observed parameter values $\eta\sub{meas}$
under the condition that the imposed constraints are fulfilled,
which are expressed by a number of constraint functions $g\sub{k}\,(\eta\sub{i})=0$.
No assumption is made, or is necessary, about the 
distribution of the true parameter values $\eta\sub{i, true}$.

However, if an ensemble of events is considered where
the distribution of a parameter $\eta\sub{true}$
is known to be Gaussian with zero mean,
then for this ensemble the choice $\eta\sub{meas} = 0$ 
also leads to a Gaussian distribution of 
$\eta\sub{meas} - \eta\sub{true}$,
and for such an ensemble it is justified 
to estimate $\eta\sub{true}$ by means of a $\chi^2$ fit.

In the case of photon radiation, the distribution of 
the unmeasured momentum $\pzgamma$ is known, though
definitely non-Gaussian.
Thus we seek a parametrization of the photon's momentum 
$\pzgamma = \pzgamma\,(\eta)$ such that the true value of $\eta$ follows 
a Gaussian distribution with mean zero and unit standard deviation 
$\delta \eta = 1$.
Then the photon will be treated as if it had a measured value of
$\eta\sub{meas}=0$. The photon will then be added to the list of fit objects
in the kinematic fit, thereby introducing an additional contribution to
the overall $\chi^2$ of $\eta^2/\delta \eta^2 = \eta^2$. 
By this procedure, the a priori knowledge of the
photon's energy spectrum (in particular the fact that it is
negligibly small in most cases) is used, and all energy and momentum
constraints can be applied.

The probability density function ${\mathcal {P}}\,(y)$ for the 
energy fraction $y = \Egamma/\Ebeam$ carried by 
initial state radiation
is well approximated by \cite{bib:isr}
\begin{equation}
  \label{eq:photonenergyspectrum}
   {\mathcal {P}}\,(y) = \beta\,y^{\beta - 1},
\end{equation}
with the exponent $\beta$ given by
\begin{equation}
  \label{eq:betavalue}
   \beta = \frac{2 \alpha}{\pi}\, \left ( \ln \frac{s}{m\sub e^2} -
   1\right ),
\end{equation}
which corresponds to $\beta = 0.1235$ for $\sqrt s = 500\,\GeV$.

Considering that an ISR photon can be emitted by either beam
leads to
\begin{equation}
	\label{eq:approx}
	{\mathcal {P}}\,(\pzgamma) = \frac {\beta} {2
        \Emax} \cdot \left | \frac{\pzgamma}{\Emax} \right |^{\beta-1},
\end{equation}
where $\Emax \le \Ebeam$ is the maximum possible photon energy.
As a consequence, the quantity $z$ given by
\begin{equation}
	z = \sign(\pzgamma)\,\left ( \frac{|\pzgamma|}{\Emax} \right ) ^\beta
        \label{eq:z}
\end{equation}
is uniformely distributed between $-1$ and $1$,
and hence
\begin{equation}
	\eta = \sqrt{2} \cdot \erf^{-1}\,(z),
        \label{eq:eta}
\end{equation}
follows a Gaussian distribution with zero mean and unit standard deviation.
Here, $\erf^{-1}\,(z)$ denotes the inverse of
the error function given by
$\erf\,(x) = \frac{2}{\sqrt{\pi}} \int\limits_0^x \E^{-t^2} \der t$.

Conversely the expressions for $z$ and $\pzgamma$ as a function of the parameter $\eta$ read
\begin{eqnarray}
  z\,(\eta) &= & \erf(\eta/\sqrt{2}) \\
  \pzgamma\,(\eta) &=& \sign(z)\,\Emax |z|^{\frac{1}{\beta}} \\
  &=& \sign(\eta)\,\Emax \, \left [ \erf(|\eta|/\sqrt{2}) \right ]^{\frac{1}{\beta}}.
  \label{eq:parametrization}
\end{eqnarray}

\subsection{Properties of the Parametrization}

\begin{figure}[ht]
	\subfigure{
		\label{fig:1a}
		\epsfig{file=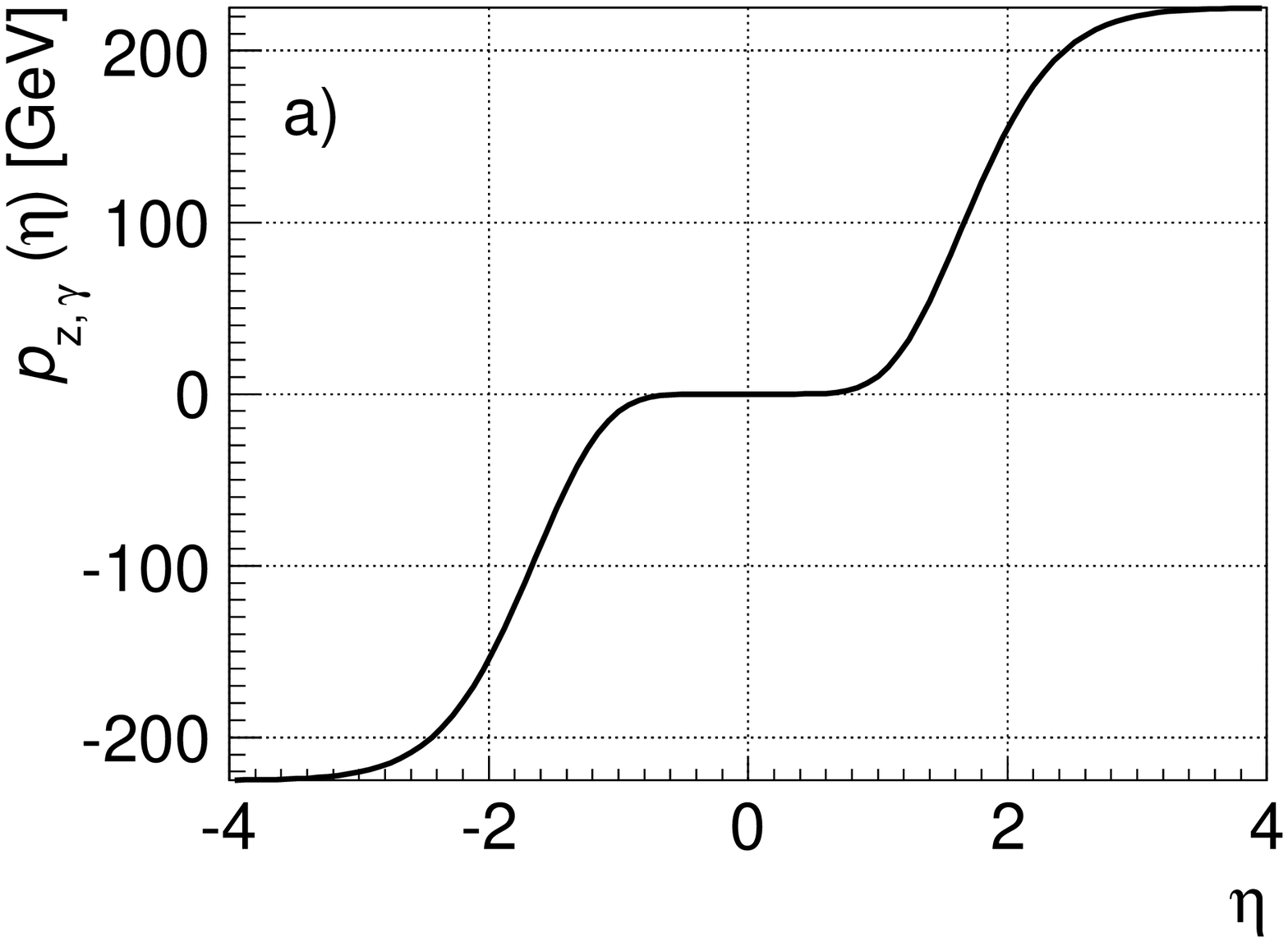,width=0.475\textwidth}
	}
	\subfigure{
		\label{fig:1b}
		\epsfig{file=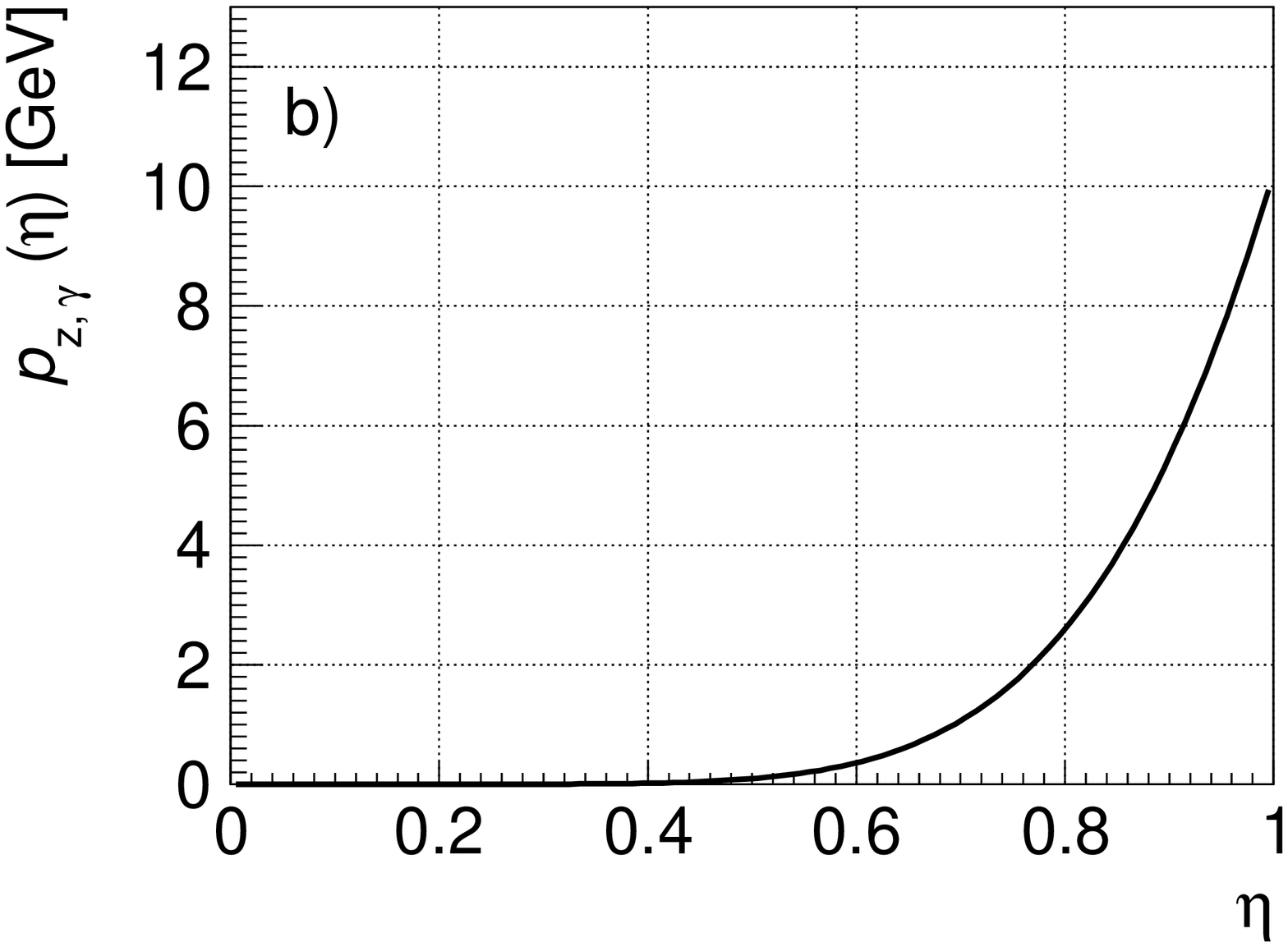,width=0.475\textwidth}
	}
	\caption{The photon's $z$-momentum $\pzgamma$ as a function of the
        fit parameter $\eta$, as given by Eq.~(\ref{eq:parametrization}),
        for $\Emax=225\,\GeV$ and $\beta=0.1235$
        in the range $|\eta|<4$ (a) and $0 < \eta < 1$ (b).
 	\label{fig:parametrization}
        }
\end{figure}

Fig.~\ref{fig:parametrization} shows a graph of $\pzgamma\,(\eta)$
for $\Emax=225\,\GeV$ and $\beta=0.1235$. The function 
has four distinct bends around $|\eta| \approx 1$ and 
$|\eta| \approx 2.5$.
It is flat around $\eta = 0$, reflecting the fact that the 
majority of ISR photons have negligible momentum;
only for $|\eta| > 0.7$ significant 
momenta above $1\,\GeV$ are predicted.

Around $\eta=0$ the value of $\pzgamma$ does not change and thus cannot influence
the global $\chi^2$ of the kinematic fit.
Therefore, the penalty term $\eta^2$
leads to a local minimum of the $\chi^2$ at this value
of $\eta$.
This will also be the global minimum if the measured
four-momenta of the final state particles are compatible with
no missing momentum from ISR. In this case,
a fit with a photon fit object has exactly the same result
as a fit without a photon.

Due to this local minimum, any minimization method based on derivatives will always
yield $\eta=0$ if this value is used as starting value in the minimization.
Therefore, to find the global minimum, in addition a different starting value must be tried,
for instance $\eta\,(-p\sub{z, miss})$ calculated from the missing $p\sub{z}$ of the event.

For $ 1\,\lapprox\, |\eta|\, \lapprox\, 2.5$, the curve rises steeply,
so that large values of missing energy and momentum from ISR and
beamstrahlung can be accomodated by the kinematic fit at a moderate penalty $\eta^2$.
Thus, for photon momenta that are large compared to the detector resolution,
the kinematic fit should find a global minimum of the $\chi^2$ close to the true
photon momentum, with a negligible bias towards low photon energies,
despite the fact that the ``measured'' value of $\eta$ 
and thus of $\pzgamma$ is set to zero.

Above $|\eta| \approx 2.5$, corresponding to $|\pzgamma|/\Emax \approx 0.9$,
the curve flattens again, so that extremely large photon momenta
are suppressed due to the fact that they are increasingly unlikely.
This region is, however, of little interest in realistic analyses.

\section{Performance Tests}
\label{sec:performance}

The method described above is applied to the process 
$e^+ e^- \to W^+ W^- \to 4\,{\mathrm{jets}}$ events.
The fraction of successful fits, the width and the shift
of the reconstructed $W^\pm$ mass peak are used 
to compare the performance of the various 
kinematic fit variants.

\subsection{Data Set}
\label{subsec:dataset}

The analysis sample was generated using 
the matrix element generator WHIZARD \cite{bib:whizard},
which takes into account all Feynman diagrams
leading to a given final state, including interference terms.
Here, the process $e^+ e^- \rightarrow u\bar{d}d\bar{u}$
is chosen, because it contains no heavy quarks in the final state, so that the
jet energy and angle measurement is not compromised by the presence of
neutrinos from semileptonic decays.

In addition to the dominant $W^\pm$ pair production,
also $Z^0$ pair production contributes to the formation
of this final state, and at a even smaller level
single boson production with subsequent
radiation off final state quarks.
Due to the inclusion of interference effects,
it is conceptually not possible to identify
events from these additional processes and
remove them.
The centre-of-mass energy is $\sqrt{s}=500\,\GeV$, and
the $W^\pm$ mass was set to $\mWgen$.

The initial state radiation is also simulated by WHIZARD, and one ISR photon per
incoming lepton is stored in the event record.
In contrast, the energy spread of the incoming beams and beamstrahlung is taken into
account in the event generation by a corresponding variation of the momenta of
the incoming leptons,
using a 
beamstrahlung spectrum that was simulated with GUINEA-PIG \cite{bib:guineapig}.
For this calculation the nominal beam parameter set of the ILC was assumed,
\cite{bib:beamparameters},
in particular an energy spread of $0.14\,\%$ and $0.07\,\%$ for the 
electron and positron beams, respectively, 
a beamstrahlung parameter $\Upsilon\sub{ave}=0.047$, a 
mean energy loss by beamstrahlung of $\delta\sub{BS} = 0.023$,
and a vertical disruption parameter of $D\sub{y}=19.1$.

Thus, the momenta of ISR photons are directly accessible in the event record,
while the combined effect of beamstrahlung and beam energy spread has to be
deduced from the total four-momentum of all final state particles.

A full simulation of the ILD detector 
\cite{bib:ild} is performed by the GEANT based simulation program 
MOKKA \cite{bib:mokka}.
In the event reconstruction,
which is implemented as part of the software package MarlinReco 
\cite{bib:marlinreco}, the tracks are matched to the calorimeter 
clusters by the Pandora particle flow algorithm \cite{bib:pandora} and the 
resulting reconstructed particles are forced into four jets by the 
Durham algorithm \cite{bib:durham}. 
Each of the four jets has to have a
 minimum energy of $E\sub{jet} > 5\,\GeV$ 
and a polar angle that fulfills
$|\cos \theta\sub{jet}|<0.9$.

The jet momentum four-vectors 
are parametrized in terms of energy $E\sub{jet}$, polar angle $\theta$
and azimuthal angle $\phi$,
with resolutions \cite{bib:jeterrors}:
\begin{eqnarray}
  \delta E\sub{jet}/E\sub{jet} &=& 32.24\,\%/\sqrt{E\sub{jet}} + 
  1.242\cdot 10^{-4}\, E\sub{jet}- 1.446\,\% \\
  \delta \theta &=& 0.03925/\sqrt{E\sub{jet}} + 0.3373 / E\sub{jet} \\ 
  \delta \phi &=& 0.05873/\sqrt{E\sub{jet}} + 0.3207 / E\sub{jet}.
\end{eqnarray}

Since the method presented here concerns radiation escaping the main
detector, a subsample of events is selected 
such that at generator level only
negligible energy from ISR is present
in the detector acceptance.
Therefore,
events are rejected with ISR photons of
energy $\Egammagen > 5\,\GeV$ and polar angles
$0.29^\circ \le \theta_\gamma\super{gen} \le 179.71^\circ$,
which corresponds to the acceptance of the beampipe calorimeter
(BeamCAL).
No cut is applied on the energy or direction of the 
beamstrahlung.

In order to investigate the influence of ISR and beamstrahlung
on the performance of the kinematic fit,
the event sample is divided into three subsamples
according to the total energy $\EISR$ of the ISR photons:
\begin{itemize}
\item A subsample with small $\EISR<5\,\GeV$, where the ISR is expected to
have a small effect and thus a kinematic fit is expected to perform well
without an additional photon,
is used to evaluate whether the addition of such a photon leads to a loss
of resolution.
\item A subsample with moderate ISR energy ($5 \le \EISR<30\,\GeV$)
  is used to evaluate a possible bias of the kinematic fit,
  and whether the addition of a photon removes this bias and increases
  the fraction of good fits.
\item A subsample with large $\EISR\ge 30\,\GeV$ serves to 
quantify how well the fit with photon
works, and whether it has any advantage over a fit where
the energy and longitudinal momentum constraints are dropped completely.
\end{itemize}

\subsection{Evaluation Method}
\label{subsec:fits}

In order to investigate the performance of the proposed method, kinematic fits 
are applied to the four jets in the events 
from  the test sample, comparing the event hypotheses ``4~jets'' ($4j$) 
and ``4~jets + 1~photon'' ($4j+\gamma$). 
Both event hypotheses are fitted with  
five constraints ($5C$-fit): conservation of energy, conservation of the three 
momentum components and equal di-jet masses. 
In addition, the events are fitted also using only the three constraints
($3C$-fit) that are not affected by the presence of photon radiation,
i.e. conservation of the transverse momentum components and the equal
mass constraint.

\begin{figure}[ht]
	\subfigure{
		\label{fig:mpu}
		\epsfig{file=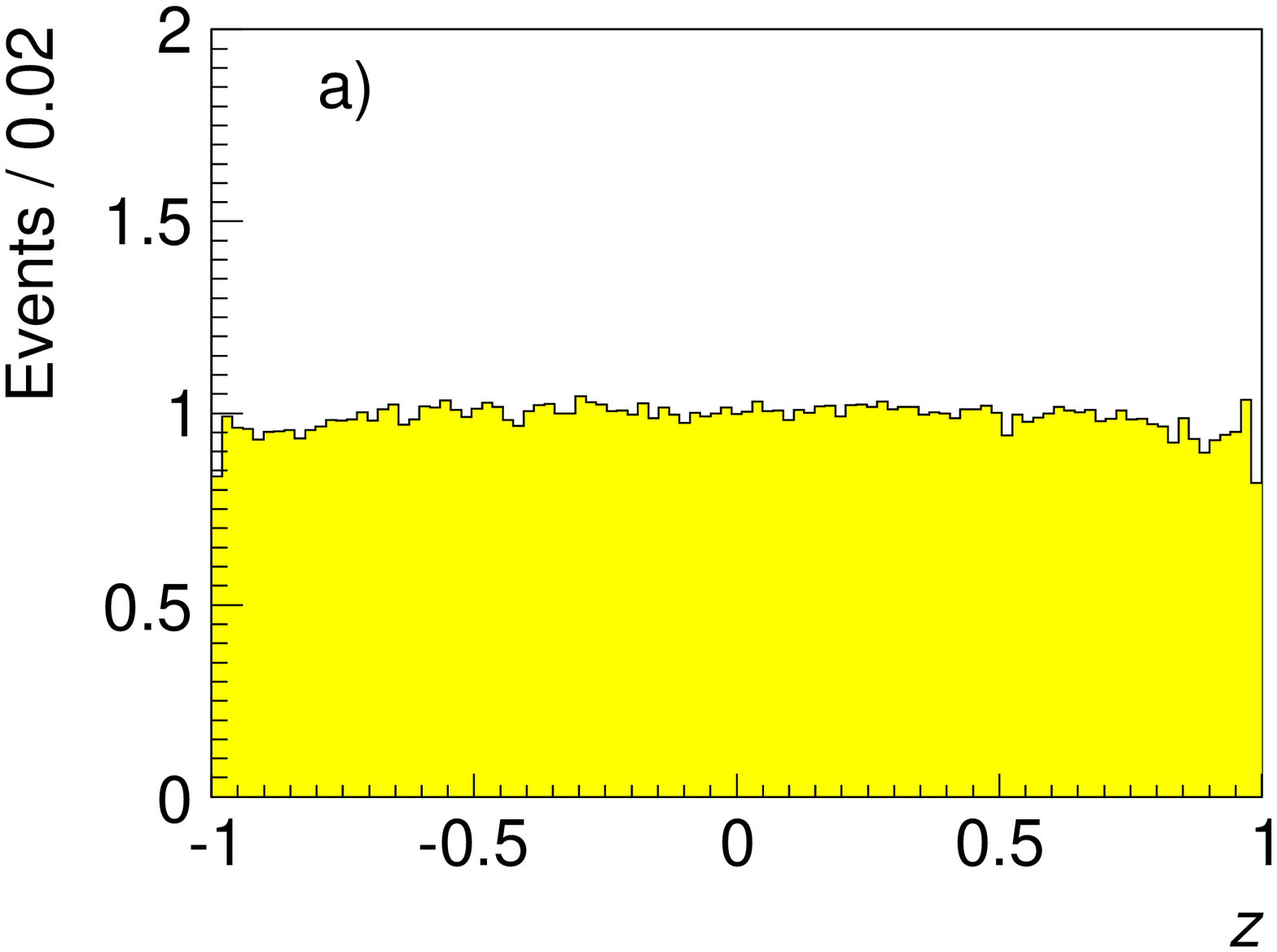,width=0.475\textwidth}
	}
	\subfigure{
		\label{fig:mpg}
		\epsfig{file=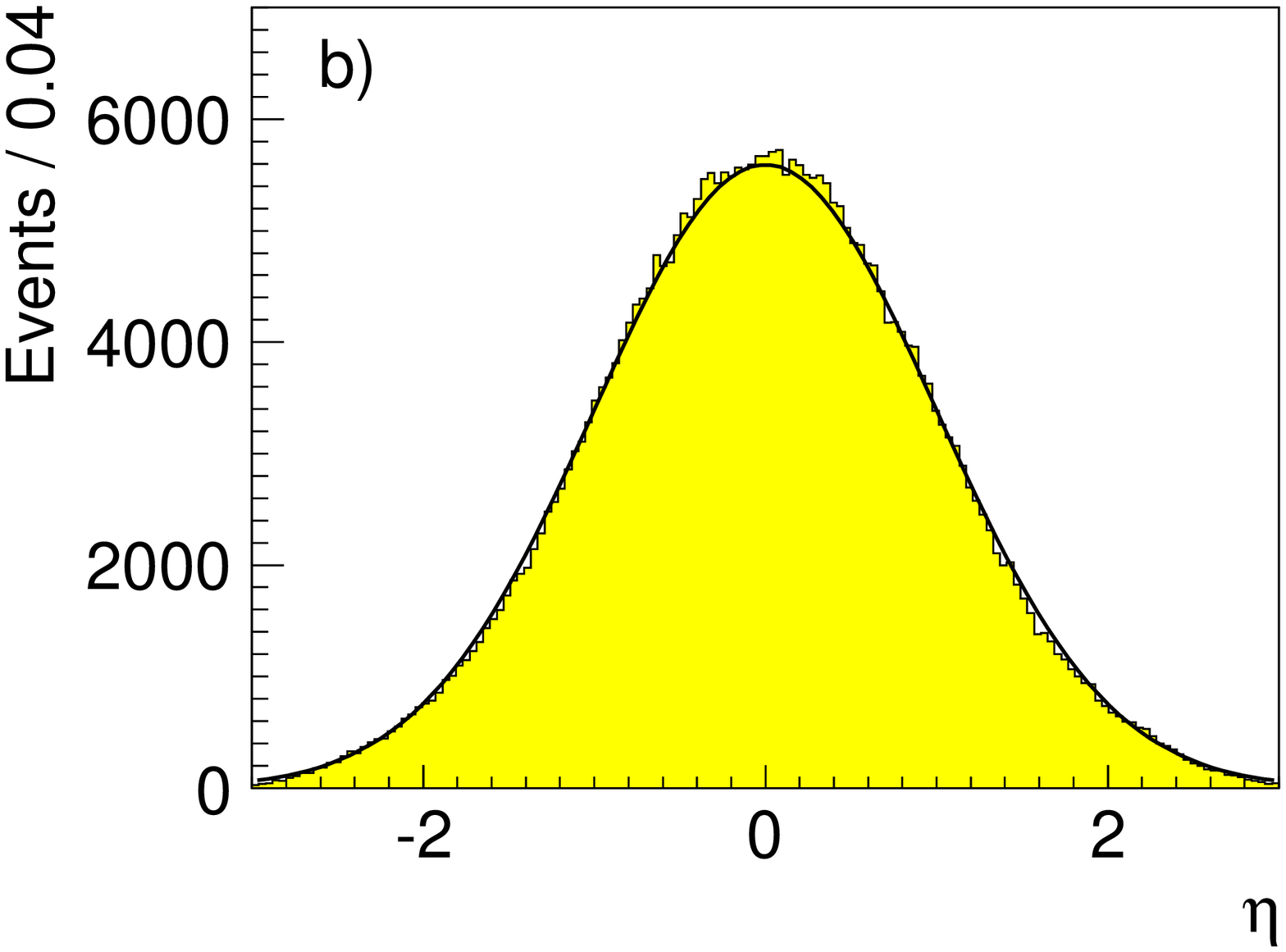,width=0.475\textwidth}
	}
	\caption{
        The parameters $z$ (a) of Eq.~(\ref{eq:z}) and $\eta$ 
        (b) of Eq.~(\ref{eq:eta}), calculated from the longitudinal
        momentum $\pzgamma$ of the most energetic ISR photon
        in the Monte Carlo sample described in the text,
        using $\Emax=225\,\GeV$ and $\beta=0.1235$.
        $z$ is expected to be uniformely distributed in
        $-1 < z < 1$,
        and $\eta$ should follow a Gaussian distribution
        with zero mean and unit standard deviation, which is shown
        for comparison in the plot.
 	\label{fig:spectrumapprox}
        }
\end{figure}

As discussed in Sect.~\ref{subsec:parametrization},
both values $\pzgamma=0$ and $\pzgamma = -p\sub{z, miss}$
are considered as starting values for the photon momentum
in the kinematic fit, and the result with the better $\chi^2$ is chosen.
Values of $\beta = 0.1235$ and $\EISR=225\,\GeV$,
which is the maximal photon energy that allows $W^\pm$ pair production,
are used in the photon parametrization
Eq.~(\ref{eq:parametrization}).
Fig.~\ref{fig:spectrumapprox}
shows the quantities $z$ and $\eta$ of Eqs.~(\ref{eq:z})
and (\ref{eq:eta}), calculated from 
$\pzgamma$ of the most energetic ISR photon in the event.
It can be seen that indeed $z$ is distributed uniformly and
$\eta$ follows a Gaussian distribution.

The sample used for the performance tests includes the effects of both
ISR and beamstrahlung.
Since the photon parametrization used here has been derived from the
ISR momentum spectrum, tests are first performed that exclude the effect of
beamstrahlung. 
The beamstrahlung is artificially ``turned off'' by using the
total generated energy and momentum of the final state particles, including the ISR
photons, in the energy and momentum constraints, rather than the nominal
values of $\sum p\sub{x, y, z} = 0$, $\sum E = \sqrt{s} = 500\,\GeV$.
Alternatively, the constraints are set to these nominal values,
so that the combined effects of ISR and beamstrahlung can be studied.

An important indicator for the performance of the various
kinematic fits considered is the fraction of good fits,
which are defined as those having a fit probability $p > 0.001$.

Due to the intrinsic widths of the $W^\pm$ and $Z^0$ bosons, 
which are not negligible 
compared to the detector resolution at an ILD-type detector,
the equal-mass constraint is only approximately fulfilled 
by the four-vectors on generator level
and therefore reduces the fraction of good fits.

The equal-mass constraint is applied mainly in order to choose the 
correct jet pairing. The fit is performed for all three possible
jet pairings, and the pairing that results in the best $\chi^2$ value
is chosen as the correct one
under the assumption that the jets stem from either a $W^{+}W^{-}$ or 
a $Z^0 Z^0$ pair.
Because of the $W^\pm$ and $Z^0$ width,
the equal-mass constraint leads to an average $\chi^2$ contribution that 
is significantly higher
than the value expected for the addition of one constraint, i.e. one
degree of freedom. As a consequence, only $55\,\%$
of the $3C$-fits have a fit probability $p>0.001$.
A more elaborate treatment of the equal-mass constraint 
that would increase this fraction is,
however, beyond the scope of the present analysis.

\begin{figure}[hbt]
	\epsfig{file=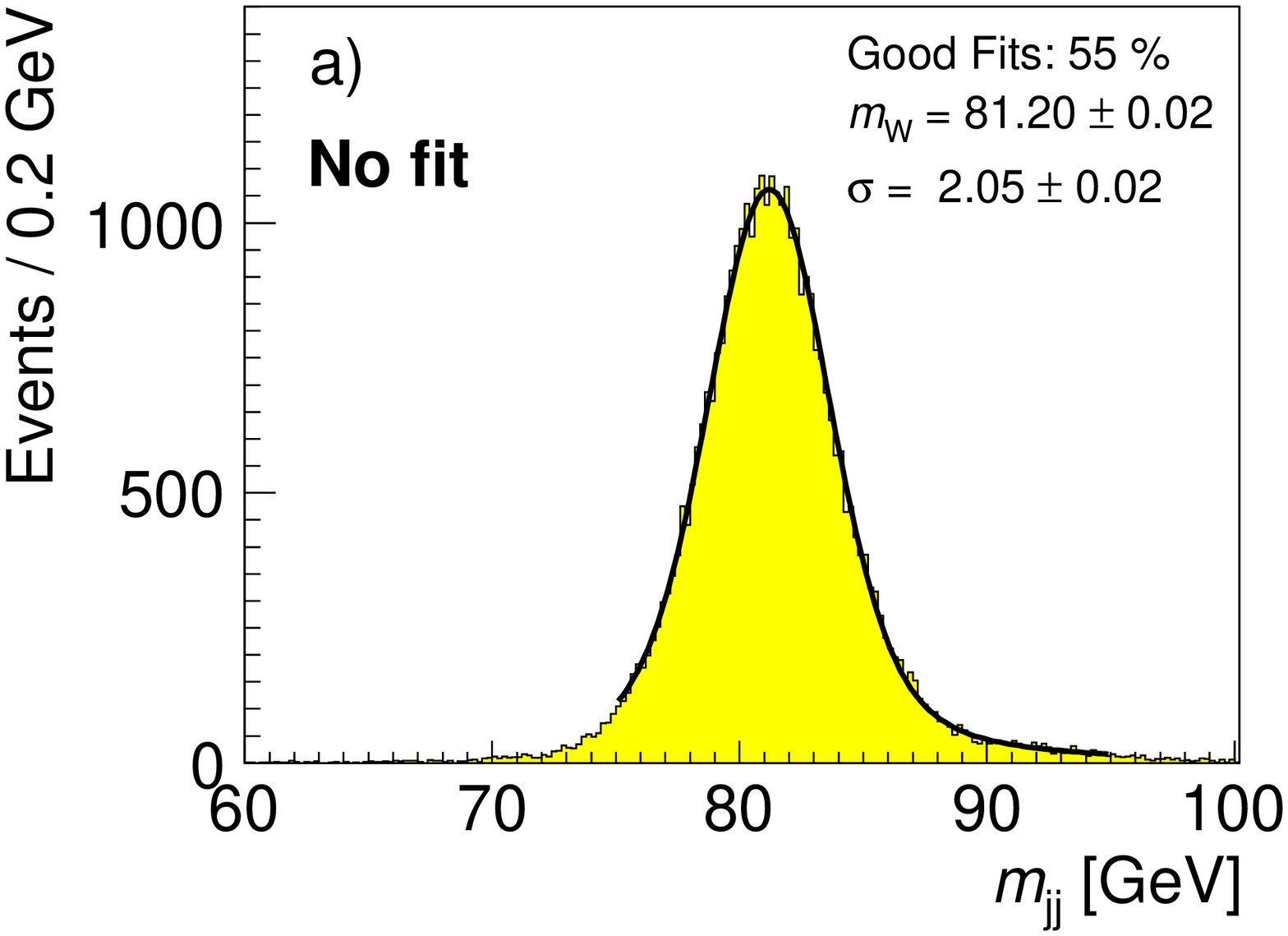,width=0.475\textwidth}
	\epsfig{file=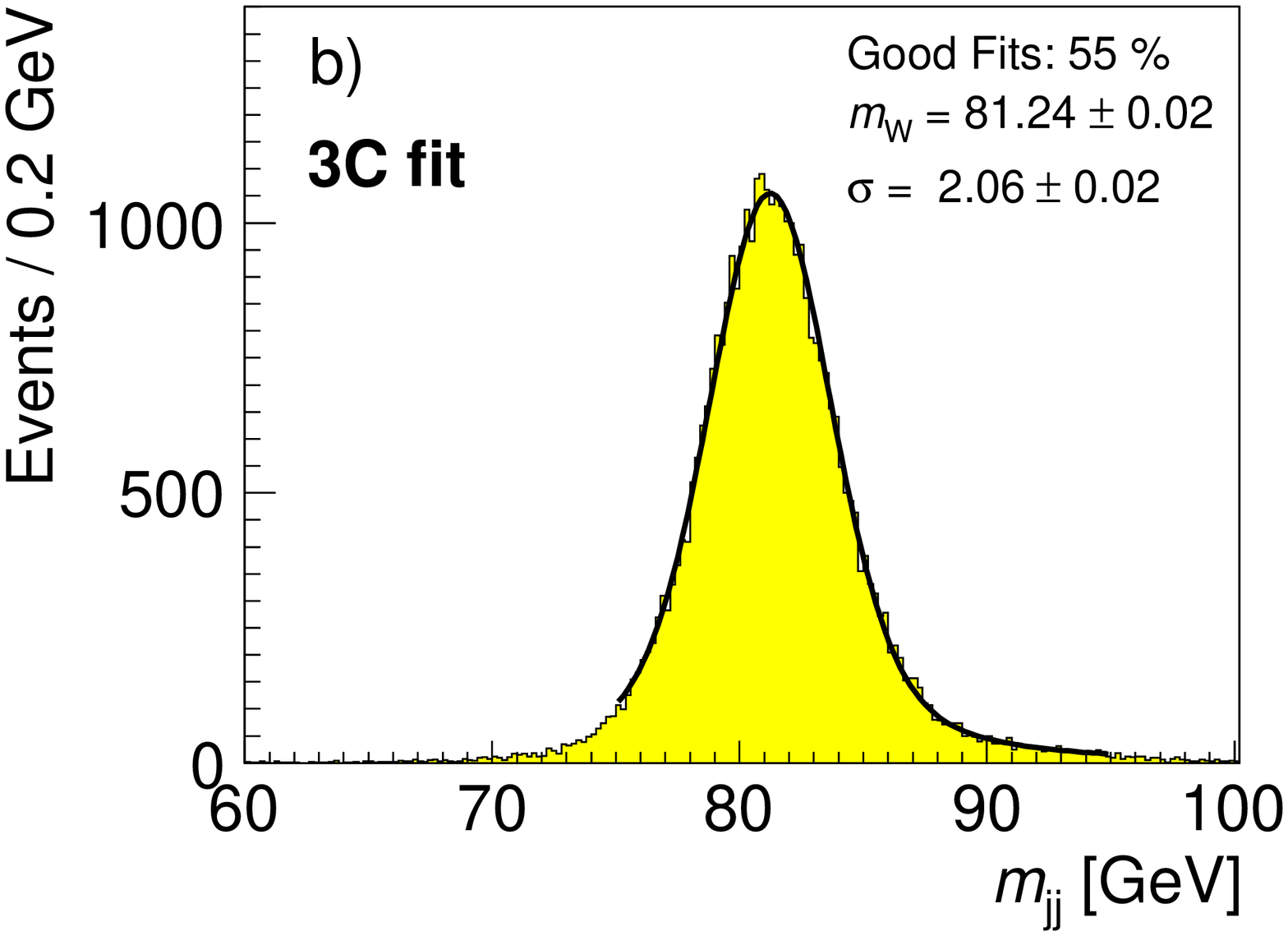,width=0.475\textwidth}\\
	\epsfig{file=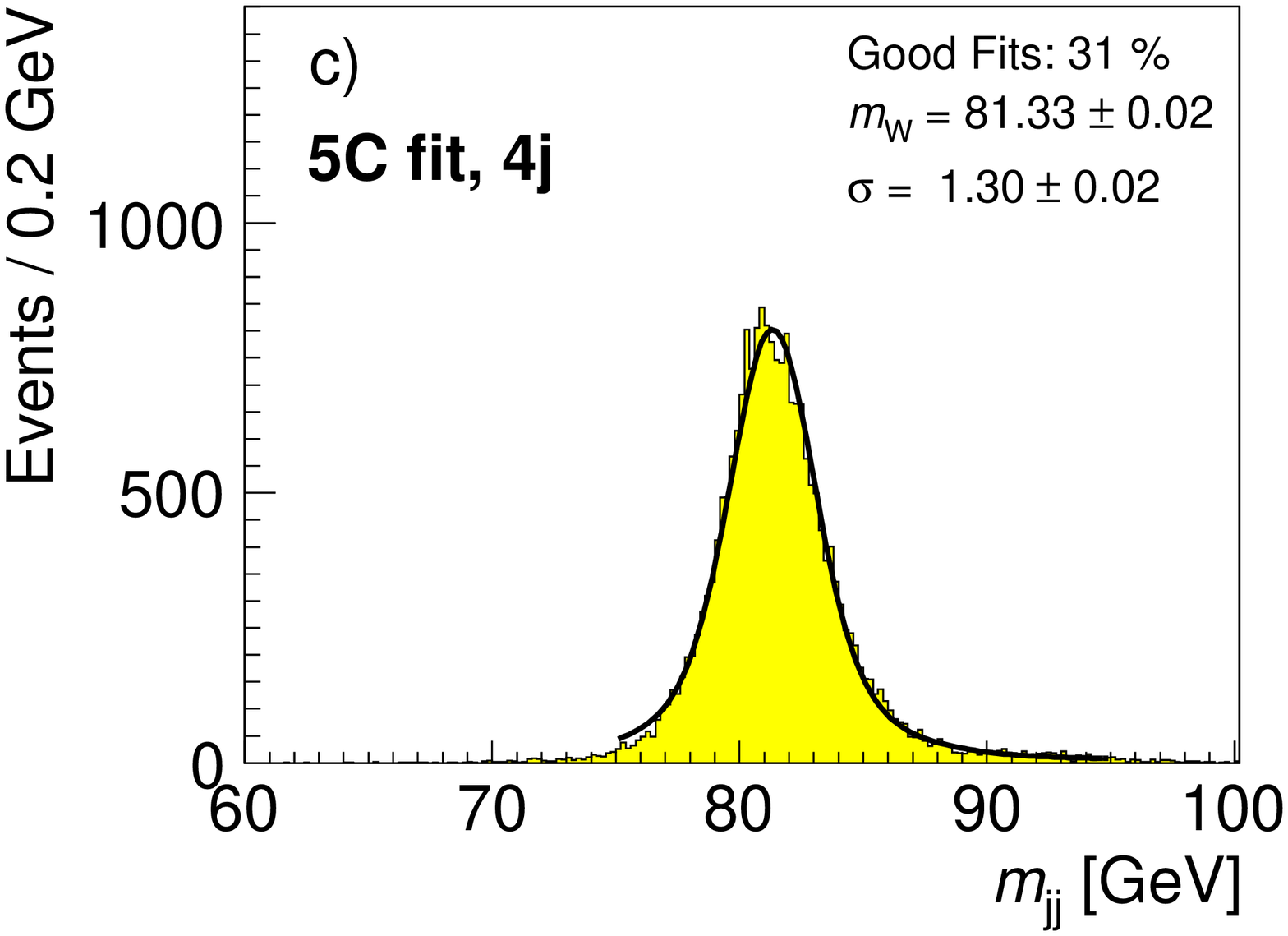,width=0.475\textwidth}
	\epsfig{file=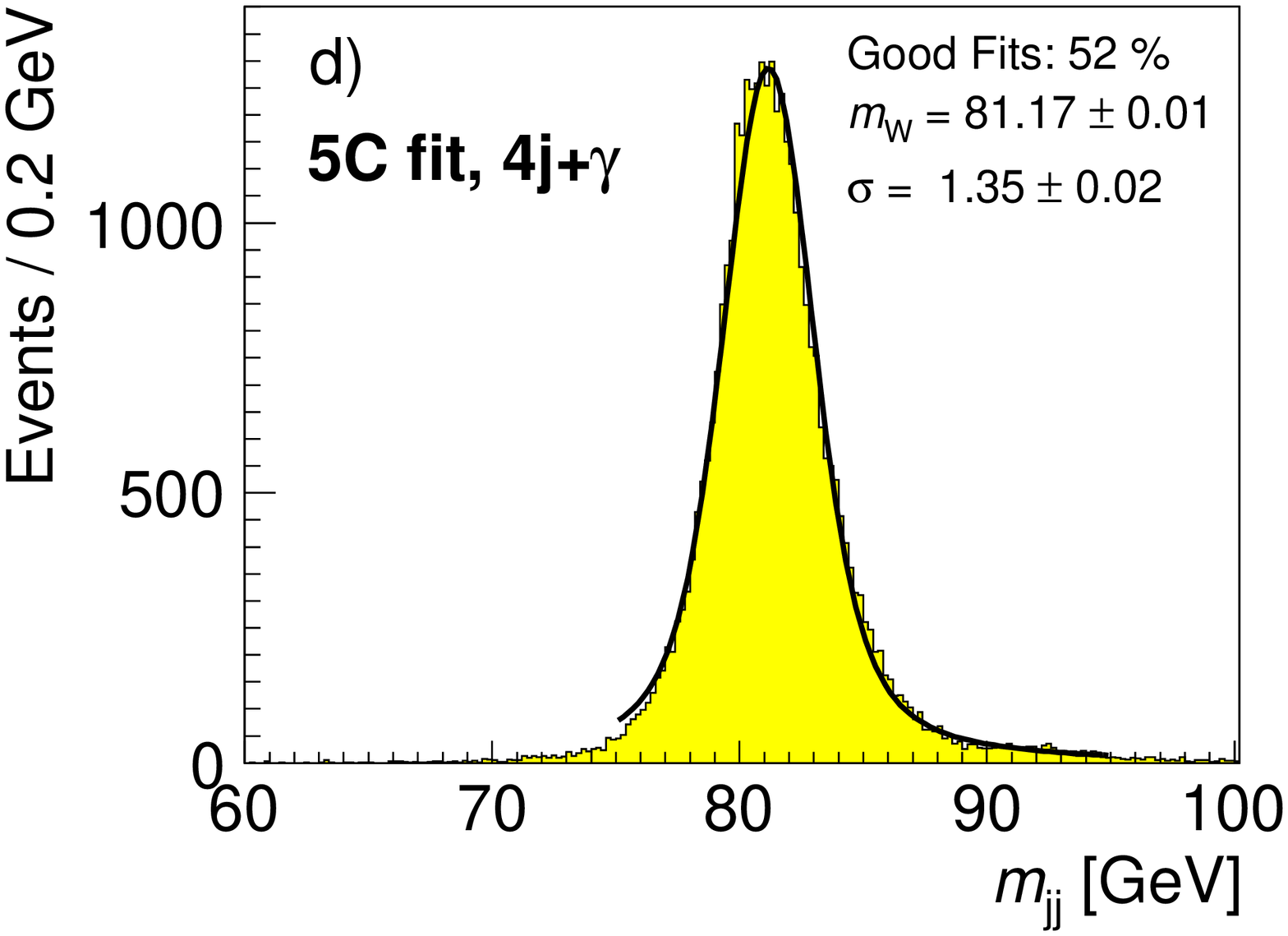,width=0.475\textwidth}
	\caption{Invariant di-jet masses $m\sub{jj}$ for the Monte Carlo 
        sample described in the text:
        a) the average of the two di-jet masses calculated 
        from the measured four--vectors, using the jet pairing
        from the $3C$ fit sample;
        b) $m\sub{jj}$ after application of the $3C$ fit;
        c) $m\sub{jj}$ for the $5C$ fit under a $4j$ hypothesis;
        d) $m\sub{jj}$ for the $5C$ fit under a $4j+\gamma$ hypothesis.
	\label{fig:dijetmasses}
        }
\end{figure}

Fig.~\ref{fig:dijetmasses} shows the the invariant di-jet 
masses before and after the kinematic fit
for the complete sample, including ISR and beamstrahlung.
A clear peak at the $W^\pm$ mass is observed, 
while the much smaller $Z^0$ mass peak appears only as 
an enhancement on the right side of the $W^\pm$ mass peak.

Imposing an equal-mass constraint leads to an implicit averaging
of the two di-jet masses in each event.
Therefore, the average di-jet mass before the fit is compared 
to the dijet masses after the various kinematic fits in Fig.~\ref{fig:dijetmasses}.

For a quantitative comparison of the different kinematic fits,
each mass distribution is fitted with an analytic function.
The $W^\pm$ mass peak is expected to follow a relativistic
Breit-Wigner distribution, folded with a Gaussian distribution that
reflects the detector resolution.
Here, the mass peaks are fitted
with a  Voigt function $V_{\sigma,\Gamma}(x)$ \cite{bib:voigtfunction},
which is the convolution of a non-relativistic Breit-Wigner (Cauchy)
distribution of width $\Gamma$ 
and a Gaussian distribution with an RMS of $\sigma$.
Thus the following function is fitted to the histograms
in the range $75 < m < 95\,\GeV$:
\begin{equation}
	f(m) = N\cdot\br{\br{1-f_Z}\cdot V_{\sigma,\Gamma_W}(m-m\sub{W})
        +f_Z\cdot V_{\sigma,\Gamma_Z}(m-m_Z)}
	\label{eq:voigt}
\end{equation}
The values for the $Z^0$ mass $m\sub{Z}=91.19\,\GeV$ and the decay widths\footnote{
The average of two independent random numbers distributed according
to a Breit-Wigner of width $\Gamma$ follows a  Breit-Wigner distribution
of the same width  $\Gamma$.
} 
$\Gamma\sub{W}=2.14\,\GeV$ and $\Gamma\sub{Z}=2.50\,\GeV$ are fixed to their
literature values \cite{bib:pdg}.
The same Gaussian width $\sigma$, reflecting the detector resolution,
is used for the $W^\pm$ and $Z^0$ mass peaks.
$N$ corresponds to the number of histogram 
entries and $f_Z$ to the fraction of $Z$-pair events. 
However, because the symmetric Voigt function does not describe the 
asymmetry of a relativistic Breit-Wigner distribution correctly, $f_Z$ 
is not an accurate estimate of the fraction of $Z^0$ events in the sample.

The parameters of interest are the Gaussian width $\sigma$ and
the difference $\Delta m\sub{W} = m\sub{W} - m\sub{W}\super{gen}$
between the fitted $W^\pm$ mass $m\sub{W}$ and 
the input $W^\pm$ mass $m\sub{W}\super{gen}$. 
 
If large amounts of energy are missing, the fitted jet energies have 
to be larger than the measured ones to fulfill energy conservation. 
Consequently,  di-jet masses are shifted to higher values
and thus a larger $\Delta m\sub{W}$ is obtained. 
Due to the imperfections of the lineshape fit,
a nonzero value of $\Delta m\sub{W}$ is to be expected,
for which a correction would be applied in a real analysis.
However, if this mass shift depends on the amount of 
energy from ISR and beamstrahlung,
it leads to a broadening of the signal and thus a loss of
resolution; in addition, systematic uncertainties arise
from the description of the ISR and in particular the beamstrahlung
energy spectrum.
Therefore, a mass shift that is independent of the amount
of energy lost to ISR and beamstrahlung is desirable.

\subsection{Results}
\label{subsec:results}

\begin{table}[tb]
\begin{tabular}{|l|l|ccc|ccc|}
	\hline
	Subsample  & Constraints, & \multicolumn{3}{|c|}{ISR only} & 
                                    \multicolumn{3}{|c|}{Full Photon Spectrum} \\
	(Fraction) & Hypothesis   & Good        & $\Delta m\sub{W}$ & $\sigma\sub{W}$ & 
                                    Good        & $\Delta m\sub{W}$ & $\sigma\sub{W}$  \\
	           &              & fits $[\%]$ & $[\GeV]$          & $[\GeV]$        & 
                                    fits $[\%]$ & $[\GeV]$          & $[\GeV]$  \\
	\hline
	\hline
	All events                   & ---             & $55\,\%$ & $+0.78$ & $2.05$ & 
                                                         $55\,\%$ & $+0.78$ & $2.05$ \\
	$(100\,\%)$                  & $3C, 4j$        & $55\,\%$ & $+0.82$ & $2.06$ & 
                                                         $55\,\%$ & $+0.82$ & $2.06$ \\
	                             & $5C, 4j$        & $42\,\%$ & $+0.67$ & $1.21$ & 
                                                         $31\,\%$ & $+0.91$ & $1.30$ \\
	                             & $5C, 4j+\gamma$ & $54\,\%$ & $+0.53$ & $1.25$ & 
                                                         $52\,\%$ & $+0.75$ & $1.35$ \\
	\hline
	$\EISR<5\,\GeV$             & ---             & $56\,\%$ & $+0.80$ & $2.04$ & 
                                                         $56\,\%$ & $+0.80$ & $2.04$ \\
	$(75\,\%)$                   & $3C, 4j$        & $56\,\%$ & $+0.85$ & $2.06$ & 
                                                         $56\,\%$ & $+0.85$ & $2.06$ \\
	                             & $5C, 4j$        & $53\,\%$ & $+0.63$ & $1.19$ & 
                                                         $40\,\%$ & $+0.86$ & $1.27$ \\
	                             & $5C, 4j+\gamma$ & $55\,\%$ & $+0.49$ & $1.24$ & 
                                                         $54\,\%$ & $+0.69$ & $1.31$ \\
	\hline
	$5 \le \EISR<30\,\GeV$      & ---             & $54\,\%$ & $+0.79$ & $2.07$ & 
                                                         $54\,\%$ & $+0.79$ & $2.07$ \\
	$(11\,\%)$                   & $3C, 4j$        & $54\,\%$ & $+0.84$ & $2.08$ & 
                                                         $54\,\%$ & $+0.84$ & $2.08$ \\
	                             & $5C, 4j$        & $15\,\%$ & $+1.68$ & $1.25$ & 
                                                         $12\,\%$ & $+2.19$ & $1.29$ \\
	                             & $5C, 4j+\gamma$ & $53\,\%$ & $+0.71$ & $1.27$ & 
                                                         $50\,\%$ & $+1.07$ & $1.51$ \\
	\hline  
	$\EISR \ge 30\,\GeV$         & ---             & $53\,\%$ & $+0.59$ & $1.99$ & 
                                                         $53\,\%$ & $+0.59$ & $1.99$ \\
	$(13\,\%)$                   & $3C, 4j$        & $53\,\%$ & $+0.66$ & $1.99$ & 
                                                         $53\,\%$ & $+0.66$ & $1.99$ \\
	                             & $5C, 4j$        & $0\,\%$  & ---     &    --- & 
                                                         $0\,\%$  & ---     &    --- \\
	                             & $5C, 4j+\gamma$ & $47\,\%$ & $+0.64$ & $1.21$ & 
                                                         $42\,\%$ & $+0.91$ & $1.38$ \\
	\hline
\end{tabular}
\caption{
  \label{tab:results}
  Results of kinematic fits under various conditions.
  ``ISR only'' refers to the case where the effect of beamstrahlung and beam energy spread is 
  removed from the fit as explained in the text, while ``Full Photon Spectrum'' includes these effects.
  For each fit variation, the fraction of good fits with fit probability $p>0.001$,
  the difference $\Delta m\sub{W}$ between the fitted and generated W mass of \mWgen, 
  and the width of the Gaussian part of the Voigt function is given.
  The rows refer to the results from averaging the measured di-jet masses without a fit
  for events where the 3C fit converges,
  the 3C fit with only transverse momentum and equal-mass constraint,
  the 5C fit under a four jet hypothesis with longitudinal momentum and energy constraints in addition,
  and the 5C fit with an additional ISR photon fit object.
  The subsamples are distinguished by the total energy $\EISR$
  of ISR photons, excluding beamstrahlung.
}
\end{table}

Tab.~\ref{tab:results} summarizes the results of our tests.
It lists the fraction of good fits, the mass shift and
the width of the Gaussian part of the Voigt function
for the complete sample, as well as the three subsamples with
different amounts of missing energy due to ISR photons.
The results are given for the average of the di-jet masses 
before a kinematic fit, using the $3C$ jet pairing,
as well as the di-jet mass after applying a $3C$ fit or
a $5C$ fit without or with an ISR photon.
The results are reported for the case where the effect from 
beamstrahlung has been excluded by adjusting the energy and momentum 
constraints (cf. Sect.~\ref{subsec:fits}), and 
for the realistic case where effects from ISR and beamstrahlung are 
fully taken into account.

\subsubsection*{Results with ISR only}

A comparison of the fit results
demonstrates the gain in resolution 
achieved by kinematic fitting: The Gaussian $\sigma$, which corresponds to
the di-jet mass resolution, is $\sigma = 2.1\,\GeV$ for the average of the 
two di-jet masses without a kinematic fit
and improves to $\sigma = 1.3 \,\GeV$ if a kinematic fit with five constraints
is used.
A fit with only three constraints does not improve the 
resolution compared to the simple averaging of the unfitted di-jet masses.

The fit with five constraints and no ISR photon cannot be applied to the
subsample with $\EISR \ge 30\,\GeV$, because fit probabilities above
the cut of $p = 0.001$ are essentially never achieved
due to the missing energy and momentum that
are are too large to be accomodated by the experimental resolution of a
few $\GeV$.
Therefore this subsample, which contains $13\,\%$ of all events, cannot be used
for an analysis.
The $5C$ fit with an ISR photon, on the other hand, achieves almost the same
performance for the two subsamples with $\EISR \ge 30\,\GeV$ and $\EISR < 5\,\GeV$
in terms of the fraction of good fits ($47\,\%$ vs. $55\,\%$)
as well as in resolution ($\sigma = 1.21\,\GeV$ vs. $1.24\,\GeV$)
with only a small additional bias in the W mass 
($\Delta m\sub{W} = 0.64\,\GeV$ vs. $0.49\,\GeV$).

The sample with moderate ISR energy  $5 \le \EISR<30\,\GeV$,
which comprises $11\,\%$ of the events,
demonstrates that the $5C$ fit without the inclusion of an ISR photon
tends to develop a mass bias. 
This is because the energy carried away by the photon
is falsely attributed to the final state jets, which increases their energy
and thus the invariant mass: The mass bias increases from
$\Delta m\sub{W} = +0.63\,\GeV$ to $+1.68\,\GeV$. At the same time, only
$15\,\%$ of the events yield a good $5C$ fit under the $4j$ hypothesis.
In contrast, the $4j+\gamma$ hypothesis shows the same performance in terms of 
fraction of good fit, mass shift and resolution as for the sample with
small missing energy.

The fact that for all fit hypotheses only about half of the events
have reasonable fit probabilities $p > 0.001$ can be mostly attributed to the 
equal-mass constraint: The resolution for the difference of
the di-jet masses is approxiately $4.1\,\GeV$ (twice the resolution for the 
di-jet mass average for the unfitted jets),
which is of similar size as the broadening\footnote{
The difference of two Breit-Wigner distributed random numbers follows itself 
a Breit-Wigner function with a width that is the sum of the two individual widths.
} of $4.3\,\GeV$ due to the intrinsic W width. 
This indicates that in a real analysis 
the na\"ive equal-mass constraint has to be modified
to take the natural $W$ width into account.
Other factors that reduce the fraction of successful fits are
events from processes other than $W/Z$ boson pair production
and the fact that the jet error parametrization employed in this 
analysis does not include the effects of parton showering.

\subsubsection*{Results with ISR and beamstrahlung}

The right-hand side of Tab.~\ref{tab:results}
shows the results for the case where the effect of both,
ISR and beamstrahlung, is considered.
Because the three subsamples are defined on the basis of
the ISR energy only, the same amount of beamstrahlung is present
in each of them.
A comparison with
the case where only the effect from ISR is considered,
demonstrates that the photon momentum parametrization
Eq.~(\ref{eq:parametrization}) derived from the ISR momentum
spectrum also works quite well in the presence of beamstrahlung,
at least at the level of beamstrahlung that is expected
for the nominal ILC parameter set.

Since beamstrahlung in the Monte Carlo simulation used for this analysis
is simulated solely through a variation of the energy of the incoming 
leptons, no transverse momentum is carried by the beamstrahlung.
Therefore the results for the $3C$ fit and the 
di-jet masses calculated without a kinematic fit do not change
when beamstrahlung effects are considered.

The performance of the $5C$ fit under the $4j$ hypothesis
is significantly reduced when beamstrahlung effects are considered
due to the larger amount of missing energy.
Overall, the fraction of good fits goes down from $42\,\%$ to
$31\,\%$. For the subsample with less than $5\,\GeV$ of
ISR energy it is reduced from $53\,\%$ to
$40\,\%$.
At the same time, the $W^\pm$ mass shift increases by approximately $0.2\,\GeV$
for the whole sample. For the subsample with medium $\EISR$, however,
the mass shift increases from $+1.68\,\GeV$ to $+2.19\,\GeV$.

On the other hand, with the $4j+\gamma$ hypothesis, the $5C$ fit performance
is much less affected by beamstrahlung effects:
The fraction of good fits stays almost constant,
and the $\sigma$ of the Gaussian width of the mass peak
increases only moderately, from $1.25\,\GeV$ to $1.35\,\GeV$ for the 
complete sample.
The mass shift increases by approximately $0.2\,\GeV$
for the full sample, which is similar to the $4j$ hypothesis.
However, for the subsample with $5 < \EISR < 30\,\GeV$
the mass shift is significantly reduced from
$+2.19$ to $+1.07\,\GeV$ by the inclusion of the photon in the fit.
The increase of the mass shift with respect to the ISR only case 
indicates that the $4j+\gamma$ hypothesis cannot fully accomodate
beamstrahlung effects,
because typically both beam particles radiate off significant energy.
This may necessitate the inclusion of 
a second photon in the fit.

\begin{figure}[hbt]
	\label{fig:m35x}
	\centering
	\subfigure{
		\epsfig{file=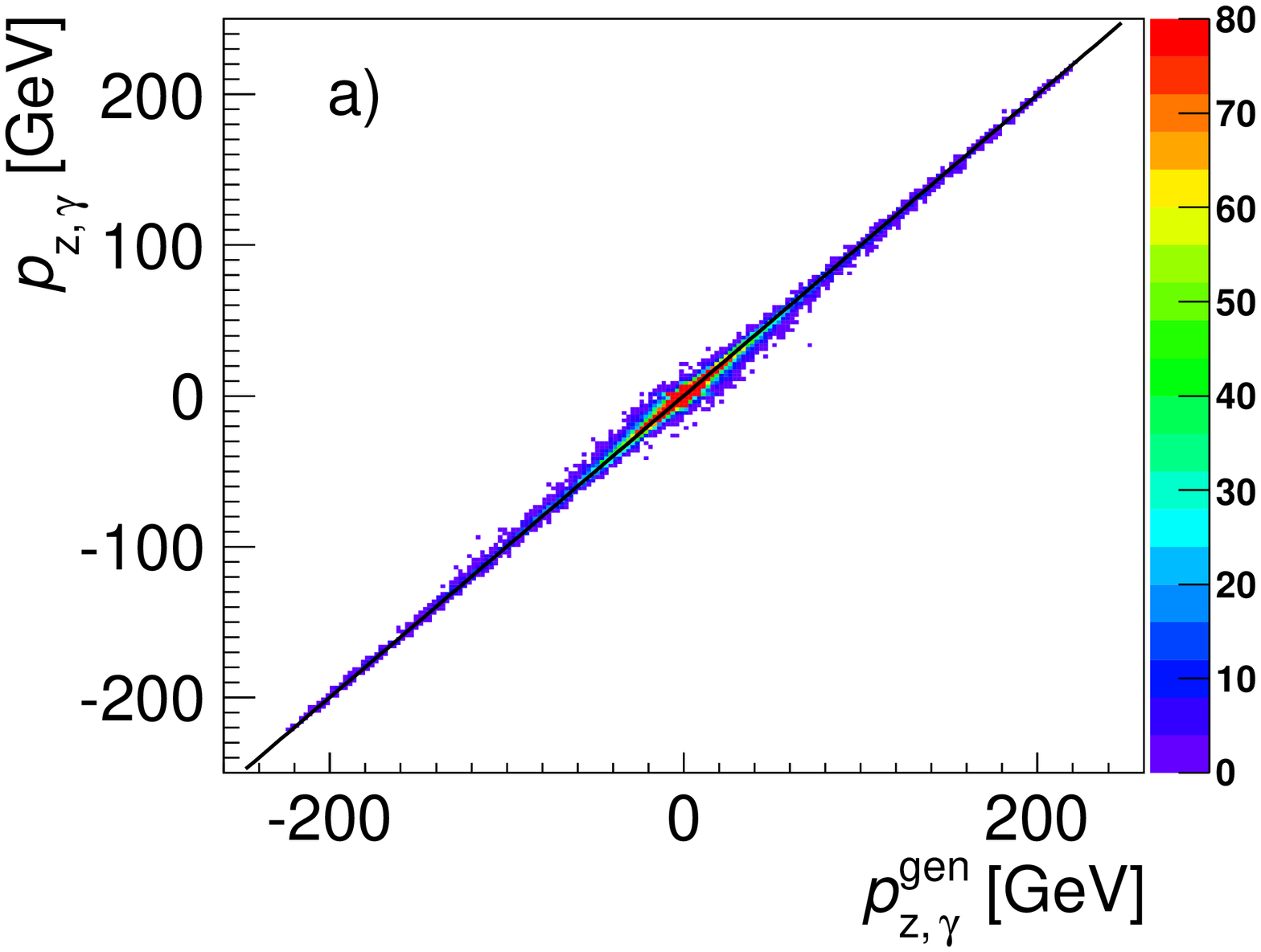,width=0.475\textwidth}
	}
	\subfigure{
		\epsfig{file=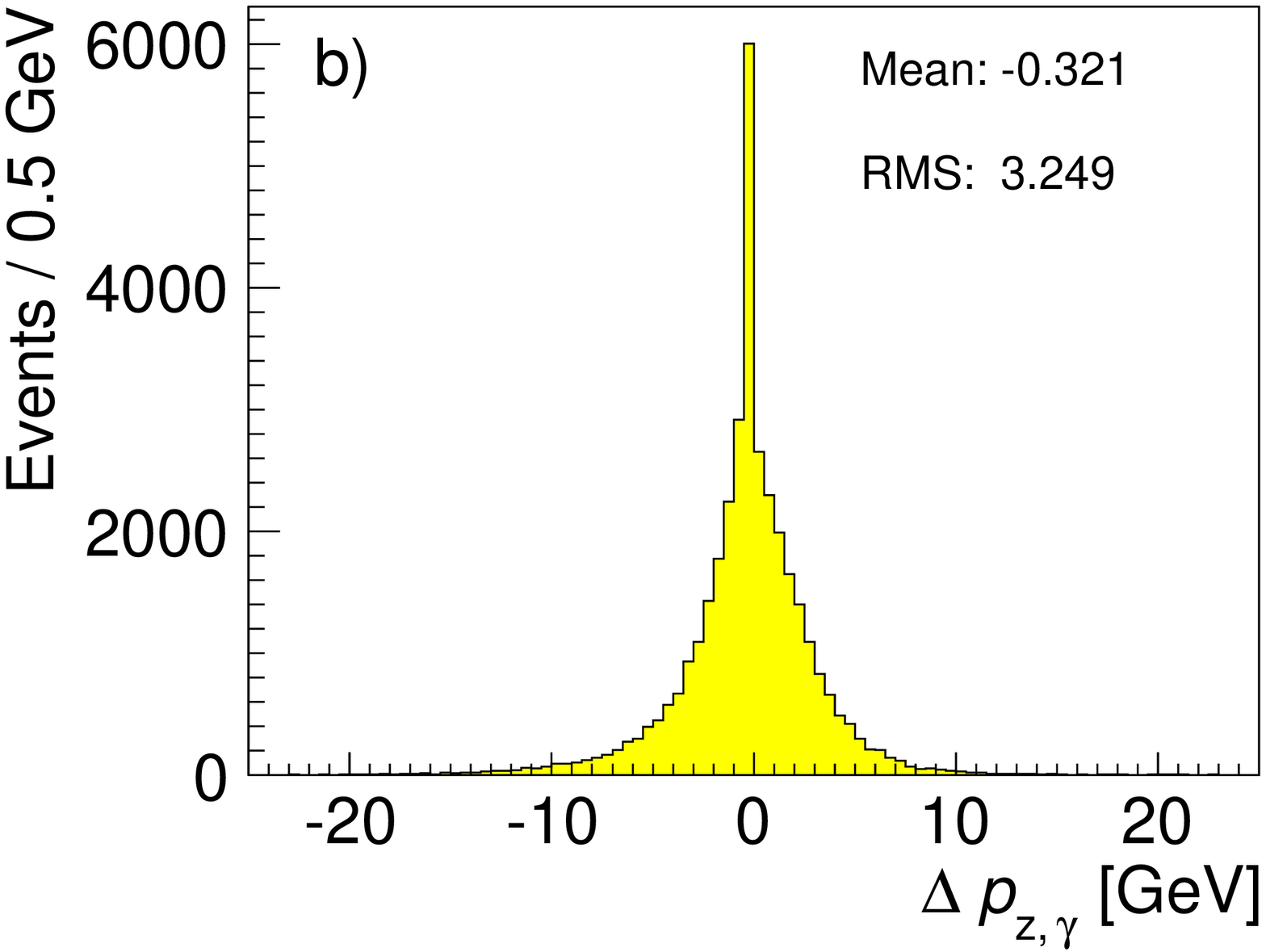,width=0.475\textwidth}
	}
	\caption{Fitted photon momentum $\pzgamma$ plotted against  
        the true momentum $\pzgammagen$ of the most energetic ISR+beamstrahlung
        photon combination in the event (a), and the difference 
        $\Delta \pzgamma = \sign (\pzgamma) \cdot (\pzgamma - \pzgammagen) $ (b).
	\label{fig:pzresoltion}
        }
\end{figure}

As a final check,
Fig.~\ref{fig:pzresoltion} shows the fitted longitudinal momentum $\pzgamma$ 
of the photon versus the generated $\pzgammagen$ of the most energetic ISR+beamstrahlung
photon pair in the event,
where the momenta of the ISR and beamstrahlung photons with either positive or
negative $\pz$ are added.
It can be seen that the fitted photon momentum $\pzgamma$
corresponds quite well to the true momentum, without any visible bias.
In particular, the fact that the photon is treated as having 
a measured $\pzgamma=0$ does not lead to a large bias towards
small values of $\pzgamma$.
This is explained by the fact that the function $\pzgamma\,(\eta)$ of
Eq.~(\ref{eq:parametrization}) rises very rapidly, as discussed in 
Sect.~\ref{subsec:parametrization}.

The right side of Fig~\ref{fig:pzresoltion} shows the difference
$\Delta \pzgamma = \sign (\pzgamma) \cdot (\pzgamma - \pzgammagen)$.
The mean $\langle \Delta \pzgamma \rangle = -0.32\,\GeV$ is small,
and negative, showing that the reconstructed $|\pzgamma|$ is slightly
smaller on average than the generated one, as expected, but that this
bias is indeed quite small. The resolution for $\pzgamma$ is found to be
$3.25\,\GeV$.


\section{Summary and Conclusions}
\label{sec:summary}

In this paper a method is proposed to take the effect 
of ISR into account in kinematic fits by introducing
a photon  that is treated as if its measured
momentum were zero. 
The longitudinal momentum $\pzgamma$ is expressed as
a function $\pzgamma\,(\eta)$ of the parameter $\eta$
such that the true
value of $\eta$ follows a normal distribution
with zero mean and unit standard deviation.

The performance of this method is evaluated using a sample
of $e^+ e^- \rightarrow u\bar{d}d\bar{u}$ events,
which is dominated by $W^+W^-$ pair production, 
at $\sqrt{s} = 500\,\GeV$.
The sample includes the effects from ISR and beamstrahlung.
It is fully simulated and reconstructed,
using the simulation for the ILD detector at the ILC.
A $5C$ kinematic fit with energy and momentum conservation constraints
and an equal-mass constraint is applied, 
and the results for the fit hypothesis with four jets and a photon
are compared to three alternatives: a $5C$ fit with
a conventional four jet hypothesis,
a $3C$ fit where the energy and longitudinal momentum constraints are 
dropped,
and the results obtained without a kinematic fit.

The $5C$ fit with the new $4j+\gamma$ hypothesis performs
as well as a $5C$ fit with a $4j$ hypothesis in terms of 
resolution, while a $3C$ is significantly worse and
does not yield any improvement over a mass reconstruction
without any kinematic fit.

For events with significant energy
from ISR photons ($5 < \EISR < 30\,\GeV$), 
the fraction of good fits with a fit probability
$p>0.001$ drops from $40\,\%$ to $12\,\%$ for a $5C$ fit
without a photon,
and goes to zero for $\EISR > 30\,\GeV$.
In addition, as the missing energy is distributed to the jets 
by such a fit,
a shift of the reconstructed di-jet masses towards larger values
is observed. 

Both problems are solved by the new $4j+\gamma$ hypothesis: 
even for large
values of $\EISR > 30\,\GeV$, 
the fraction of good fits 
and the di-jet mass resolution are similar to the values obtained
at $\EISR < 5\,\GeV$,
while the mass shift remains small.

In short, under the $4j+\gamma$ hypothesis, 
a $5C$ fit achieves the same resolution
as with a conventional $4j$ fit hypothesis, 
but independent of the amount
of ISR energy, without developing a mass bias,
and with a similar fraction of good fits as a $3C$ fit.

Although the parametrization $\pzgamma\,(\eta)$ was
developed using the momentum spectrum of ISR photons,
the method also performs well in the presence of
beamstrahlung, at least at the moderate level
expected for the nominal parameter set of the ILC.

In a future development the parametrization could be adapted
to include beamstrahlung effects. This may be necessary
in scenarios with enhanced beamstrahlung, such as the ``low power''
parameter set proposed for the ILC, or at CLIC.
We expect that under such conditions 
the addition of a second photon in the fit
would become necessary in order to take into account
the energy loss suffered by both beam particles.

\section*{Acknowledgements}
\addcontentsline{toc}{section}{Acknowledgements}

We would like to thank the ILD simulation production team, in particular F. Gaede, 
S. Aplin, J. Engels and I. Marchesini, for the production of the 
samples of events used in this work,
and T. Barklow for producing the generated
input files. 

We acknowledge the support of the DFG through the SFB (grant SFB 676/1-2006)
and the Emmy-Noether program (grant LI-1560/1-1).

\clearpage

%
%

\addcontentsline{toc}{section}{Figures}
\addcontentsline{toc}{subsection}{Fig. 1}

\addcontentsline{toc}{section}{Figures}
\addcontentsline{toc}{subsection}{Fig. 2}

\addcontentsline{toc}{subsection}{Fig. 3}

\addcontentsline{toc}{subsection}{Fig. 4}

\end{document}